\documentclass[
  reprint,
  aps,
  prx,
  superscriptaddress,
  amsmath,amssymb
]{revtex4-2}

\usepackage{graphicx}
\usepackage{tikz}
\usepackage{dcolumn}
\usepackage{bm}
\usepackage{hyperref}

\usepackage[caption=false]{subfig}
\usepackage{physics}
\usepackage{booktabs}
\usepackage{siunitx}
\usepackage[inline]{enumitem}
\usepackage[percent]{overpic}
\usepackage{soul}
\usepackage{xcolor}

\newcommand{\muon}{\ensuremath{\mu^{-}}~}
\newcommand{\electron}{\ensuremath{\mathrm{e}^{-}}}
\newcommand{\hole}{\ensuremath{\mathrm{h}^{+}}}
\newcommand{\eh}{\electron\!/\,\hole~}

\definecolor{brickred}{RGB}{178,34,34}

\begin{document}


\title{\textbf{Characterizing Quantum Error Correction Performance of Radiation-induced Errors} 
}%

\author{P.\ G.\ Baity}
\affiliation{Computing and Data Sciences, Brookhaven National Laboratory, Upton, NY 11973, USA}

\author{A.\ K.\ Nayak}
\affiliation{Department of Electrical and Computer Engineering, University of Illinois Urbana-Champaign, Urbana, IL, USA}


\author{L.\ R.\ Varshney}
\affiliation{Computing and Data Sciences, Brookhaven National Laboratory, Upton, NY 11973, USA}
\affiliation{AI Innovation Institute, Stony Brook University, Stony Brook, NY 11790, USA}

\author{N.\ Jeon}
\affiliation{Department of Electrical and Computer Engineering, Texas A\&M University, College Station, TX 77843, USA}

\author{B.-J.\ Yoon}
\affiliation{Computing and Data Sciences, Brookhaven National Laboratory, Upton, NY 11973, USA}
\affiliation{Department of Electrical and Computer Engineering, Texas A\&M University, College Station, TX 77843, USA}

\author{P.\ J.\ Love}
\affiliation{Department of Physics and Astronomy, Tufts University, Medford, MA 02155, USA}
\affiliation{Computing and Data Sciences, Brookhaven National Laboratory, Upton, NY 11973, USA}

\author{A.\ Hoisie}
\affiliation{Computing and Data Sciences, Brookhaven National Laboratory, Upton, NY 11973, USA}

\date{\today}

\begin{abstract}
Radiation impacts are a current challenge with computing on superconducting-based quantum devices because they can lead to widespread correlated errors across the device. Such errors can be problematic for quantum error correction (QEC) codes, which are generally designed to correct independent errors. To address this, we have developed a computational model to simulate the effects of radiation impacts on QEC performance. This is achieved by building from recently developed models of quasiparticle density, mapping radiation-induced qubit error rates onto a quantum error channel and simulation of a simple surface code. We also provide a performance metric to quantify the resilience of a QEC code to radiation impacts. Additionally, we sweep various parameters of chip design to test mitigation strategies for improved QEC performance. Our model approach is holistic, allowing for modular performance testing of error mitigation strategies and chip and code designs.
\end{abstract}

\maketitle


\section{\label{sec:Intro} Introduction}

The current generation of superconducting quantum processors are plagued by several sources of physical qubit noise \cite{Krantz2019}. While many studies focus on the ubiquitous two-level system noise \cite{Mueller2019,McRae2020} or electronic noise \cite{Chow2010,Hyyppa2024}, recent works \cite{Vepsalainen2020, Wilen2021, Iaia2022, Harrington2025, Larson2025} have shown that radiation impacts on quantum devices cause distinctive, correlated noise profiles in which data on neighboring qubits simultaneously decay due to the generation of superconducting quasiparticles (QPs). This spatially correlated noise is especially problematic for quantum error correction (QEC), where codes typically are designed to handle only a finite number of single-qubit errors \cite{Wilen2021,Fowler2012,Cleland2022}. Several mitigation strategies, such as phonon downconversion \cite{Martinis2021,Iaia2022} and superconducting gap engineering \cite{McEwen2024}, have been developed to address QP errors. Recently, a Monte-Carlo-based approach was created \cite{Yelton2024} to model the effects of such mitigation strategies by simulating the QP density evolution following the impact of a $\gamma$ ray or cosmic ray \muon. However, to date, this model has only been applied to test devices for verification purposes.

General studies of surface codes assume a uniform or average error rate $p_{e}$ for individual qubits with logical error probability $p_L\propto p_{e}^{(d+1)/2}$, where $d$ is the distance of the code \cite{Dennis2002,Fowler2012,Cleland2022,Acharya2025}. However, in the context of radiation impacts, error rates are expected to be both spatially and temporally correlated across the sequential QEC cycles \cite{Wilen2021}. Codes, such as surface codes that detect the parity among nearest neighbors, may be highly susceptible to uncorrectable errors or miscorrection. Studies on the failure mechanisms within running QEC codes during radiation impact events remain limited \cite{Xu2022,Ousmane2023,Vallero2024,Chadwick2024,Tan2024}, and qubit error rates are approximated using phenomenological noise models with radial symmetry from the impact site. In practice, however, a quantitative model of the noise may provide more accurate predictions in the context of phonon caustics \cite{Kelsey2023,Hernandez2025}, material dependence \cite{Place2021,Murray2021}, or the implementation of physical error mitigation strategies \cite{Martinis2021,Iaia2022}. To more completely elucidate QEC failure mechanisms in the context of radiation impacts, we develop and introduce a new computational model for QEC efficacy of radiation-induced errors. This model takes advantage of recent developments in the modeling of qubit error rates following radiation impacts \cite{Yelton2024}. We demonstrate the model by characterizing the efficacy of phonon downconversion \cite{Iaia2022} when implemented on a 17-qubit, transmon-based quantum processing unit (QPU) performing a [[9,1,3]] surface code.

\section{\label{sec:QPModel} Radiation Impact Model}
\begin{figure}
  \subfloat{
    \includegraphics[width=0.95\linewidth]{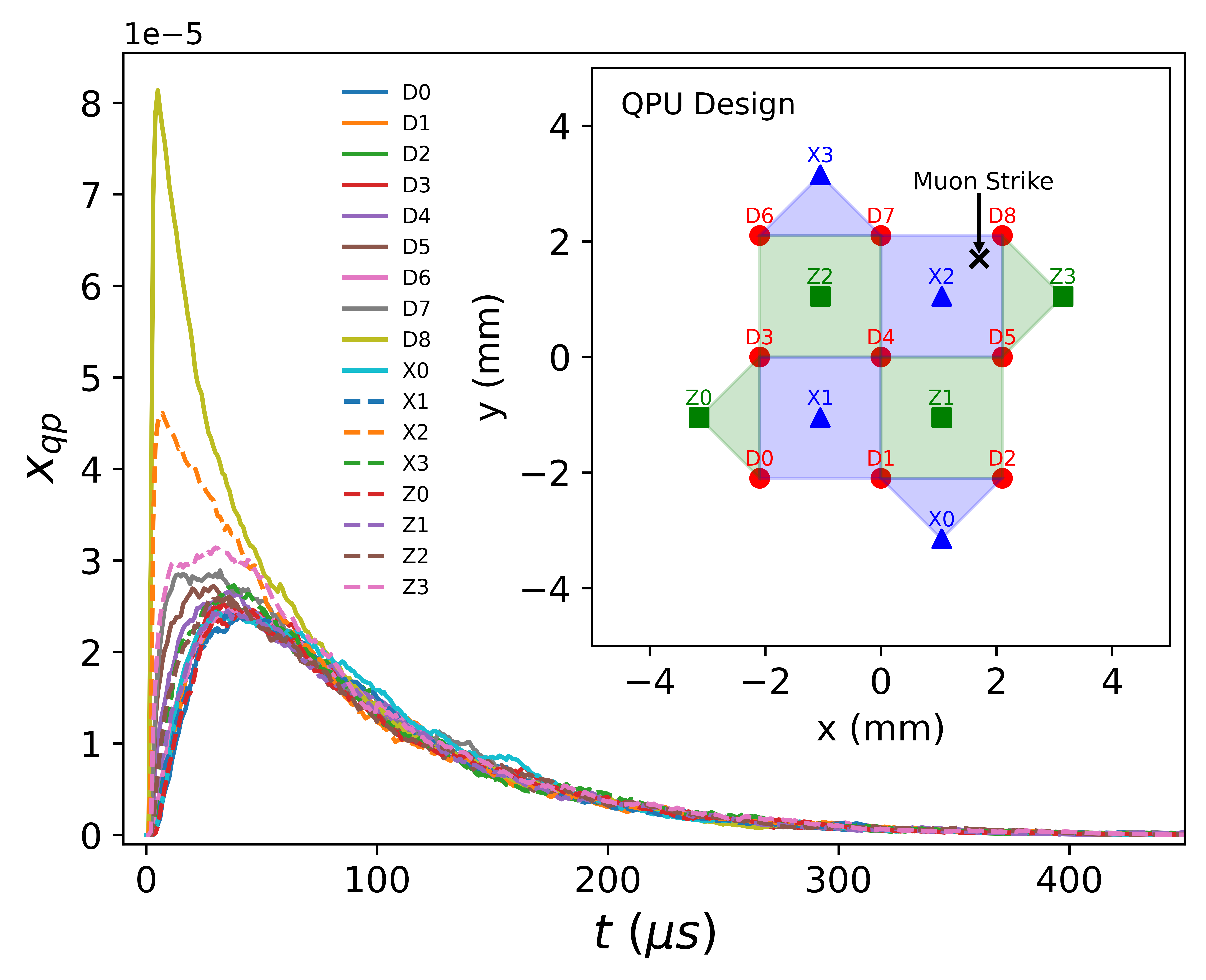}
    }
    \put(-245,170){\bfseries (a)}
  
  \subfloat{
    \includegraphics[width=0.95\linewidth]{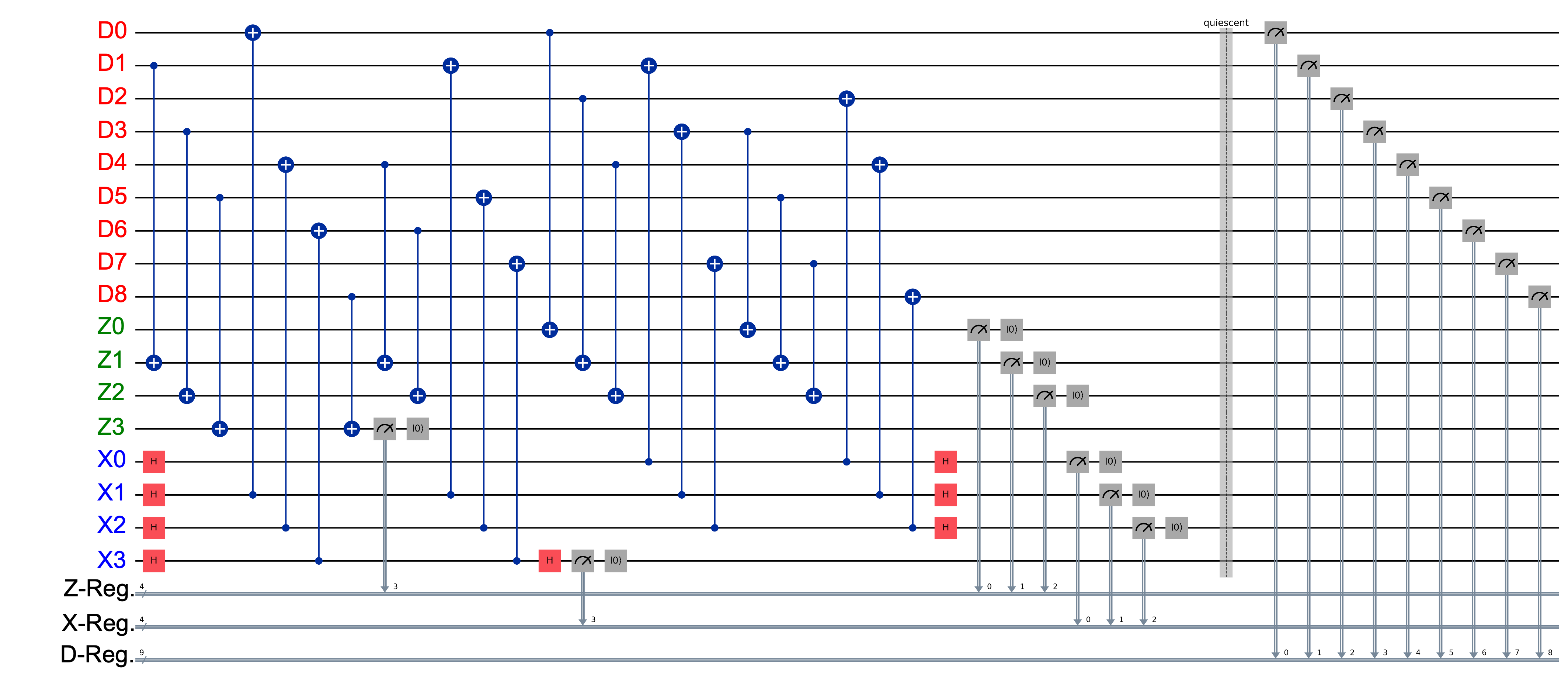}
    }
    \put(-245,90){\bfseries (b)}

  \caption{(a) The normalized quasiparticle density $x_{qp}$ for a 17-qubit QPU following a muon strike. These curves are generated using the Geant4/G4CMP simulation and ODE as described in Ref.~\onlinecite{Yelton2024} and App.~\ref{app:QPM}. The corresponding QPU design and muon strike location are shown in the inset. The device is modeled with Nb/Al superconducting material platform fabricated on $10\times10\times0.525$~mm$^3$ Si die. No Cu is modeled on the flipside of the chip.
  (inset) The QPU layout for the rotated [[9,1,3]] surface code: 17 qubits are arranged into nine data (red), four $X$-parity (blue), and four $Z$-parity (green) ancilla qubits for syndrome readout. The stabilizer plaquettes are shaded as blue and green squares and triangles to help visualize which qubits are checked by which stabilizer measurements.
  (b) The circuit diagram for the stabilizer rounds consisting of Hadamard (H) and controlled not (CX) gates. The qubit registers are labeled in accordance to those in (a). At the end of the stabilizer rounds, the syndromes are measured out to the classical register. The syndrome measurement projects the data into a quiescent state \cite{Cleland2022}, which can be saved for future use before the data readout.\label{fig:circuit}}
\end{figure}

Our approach uses a combination of computational methods to model the failure rates for QEC. The physical model for radiation impacts and QP poisoning of superconducting devices is derived from the results of Yelton et al. \cite{Yelton2024}, which uses the Monte-Carlo-based simulation packages Geant4 \cite{Agostinelli2003,Allison2006,Allison2016} and G4CMP \cite{Kelsey2023} for high-energy and condensed matter particle tracking, respectively. These packages can be used together to model the generation of \eh pairs, phonons, and QPs in planar superconducting devices. The output of these simulations is the generation term for superconducting QPs, $g_{qp}(t)$, which is used to derive the time evolution of the normalized QP density $x_{qp}$ (see Ref.~\onlinecite{Yelton2024} and App.~\ref{app:QPM}) using the time-dependent ordinary differential equation (ODE) for QP recombination and decay \cite{RothwarfTaylor1967,Wang2014,Yelton2024}. Figure~\ref{fig:circuit}(a) shows an example $x_{qp}(t)$ curve set for a 17-qubit QPU following a muon strike.

An increase in $x_{qp}$ decreases the relaxation times of the transmons linearly \cite{Catelani2012,Wang2014,Iaia2022,Yelton2024}, and the dynamical changes to the relaxation time $T_1(x_{qp})$ can be calculated following an event as
\begin{equation}
    \frac{1}{T_1}= \frac{1}{T_1^b} + \frac{x_{qp}}{\pi}\sqrt{\frac{2\omega_{01}\Delta_{Al}}{\hbar}},
\end{equation}
where $T_1^b$ is the baseline relaxation time, $\omega_{01}$ is the qubit transition frequency, and $\Delta_{Al}$ is the superconducting gap of the Al-based Josephson junction. The qubit parameters are uniform for all 17 transmons and are listed in Table~\ref{tab:transmon_params}. Moreover, while previous works have focused primarily on the relaxation time $T_1$, the quantum noise channel used in Section~\ref{sec:QECModel} also requires the dephasing rate $\Gamma_\phi$ or equivalently $T_2$. The $x_{qp}$-dependence for $\Gamma_\phi$ is derived in Appendix~\ref{app:T1T2} and depends on the transmon capacitive energy $E_C$:
\begin{equation}
    \Gamma_\phi(x_{qp})=\Gamma_\phi^b + \frac{E_C x_{qp}^2}{2 \pi^2 \hbar}e^{\tfrac{1}{2}W_0\!\bigl(4\pi/x_{qp}^2\bigr)}.
\end{equation}
Here, $\Gamma_\phi^b$ is the baseline dephasing rate, and $W_0(\cdot)$ is the principal branch of the Lambert function (i.e.,~the product logarithm function). However, as described in \cite{Catelani2012}, this dephasing rate is small compared to the relaxation rate. As a result, our analysis of QEC performance focuses on bitflip errors for $Z$-basis measurements and data correction.

\begin{table}
\begin{tabular}{ cc } 
\multicolumn{2}{c}{\textbf{Transmon Parameters}} \\
\midrule
$\omega_{01}/2\pi$  & \SI{5.000}{\giga\hertz} \\
$E_C/h$   & \SI{0.400}{\giga\hertz} \\
$E_J/h$   & \SI{9.234}{\giga\hertz} \\
$E_J/E_C$ & 23 \\
$T_1^b$   & \SI{100}{\micro\second} \\
$T_2^b$   & \SI{200}{\micro\second} \\
\midrule
\multicolumn{2}{c}{\textbf{Gate Times}} \\
\midrule
CX & 40 ns \\
H & 30 ns \\
Readout & 140 ns \\
\bottomrule
\end{tabular}
\caption{Transmon and gate parameters assumed in the simulation. These values are assigned to every qubit in the circuit. $T_1^b$ and $T_2^b$ are the baseline relaxation times for the qubits prior to any quasiparticle generation. $E_C$ and $E_J$ are the transmon capacitive and Josephson energies, respectively, and $\omega_{01}$ is the transition frequency.\label{tab:transmon_params}}
\end{table}

\section{\label{sec:QECModel} Surface Code and Error Model}
While Geant4 and G4CMP simulations provide the necessary framework to model the physical response for a \muon or $\gamma$ ray strike on the QPU, the QEC code is simulated in either Qiskit and the relevant subpackages Qiskit-Aer and Qiskit-QEC \cite{qiskit2024} or Stim \cite{Gidney2021}. Figure \ref{fig:circuit} shows the stabilizer circuit and qubit layout for the example [[9,1,3]] rotated surface code circuit with nine data qubits and eight ancilla qubits. In the surface code operation, this stabilizer circuit and corresponding syndrome measurements are repeated in consecutive cycles \cite{Cleland2022,Acharya2025}, and errors are decoded \cite{Riesebos2017,Chamberland2018,Higgott2022}. To model the dynamical qubit error rates during the simulation, the noise model is applied in a quasi-equilibrium approximation. $T_1$ and $T_2$ are evaluated with time step resolution $\Delta t$ equal to the surface code cycle time $\tau_c$, and a per cycle noise model is constructed corresponding to the instantaneous values of $T_1$ and $T_2$. This noise model is then applied to each gate within the cycle. In Qiskit, this is accomplished by passing the state vector between a series of artificial backends, each with their own noise model. In the demonstrated surface code, $\Delta t =\tau_c = 1$~$\mu$s approximately agrees with recent implementation \cite{Acharya2025} of the surface code. Because $T_1$ recovery occurs over the order of $\gtrsim 300$~$\mu$s, the dynamics are generally slow enough to accommodate the quasi-equilibrium condition for each cycle.

The cycle-dependent noise model itself is generated using a generalized amplitude damping (GAD) quantum error channel for thermal relaxation and dephasing (see Qiskit-Aer documentation \cite{qiskit-aer-noise-errors}). 
The Choi representation for this channel is:
\begin{widetext}
\begin{equation*}
\Lambda=
\begin{pmatrix}
1 - p_1(1-e^{-t_g/T_1}) & 0 & 0 & e^{-t_g/T_2} \\
0 & p_1(1-e^{-t_g/T_1}) & 0 & 0 \\
0 & 0 & p_0(1-e^{-t_g/T_1}) & 0 \\
e^{-t_g/T_2} & 0 & 0 & 1 - p_0(1-e^{-t_g/T_1})
\end{pmatrix},
\label{eq:GAD}
\end{equation*}
\end{widetext}
where $p_0$ and $p_1$ are the thermal populations of $\ket{0}$ and $\ket{1}$, respectively, and $t_g$ is the gate time. Typically, $p_0$ and $p_1$ are calculated using Maxwell-Boltzmann statistics \cite{Jin2015}, where for sufficiently cold systems, $p_0\approx1$ and $p_1\approx0$. In the context of radiation impacts, the system may be considered in terms of a quasi-equilibrium effective temperature $T_{eff}$ with increased excited state populations (see App.~\ref{app:QPM}). However, experiments \cite{McEwen2022} have shown an asymmetry in rates of $\ket{1}\rightarrow\ket{0}$ and $\ket{0}\rightarrow\ket{1}$ during radiation impacts, indicating an atypical thermal response. Although QPs are generated with densities corresponding to higher temperatures ($T_{eff}(x_{qp})\sim 250$~mK) in quasi-equilibrium, the lack of observed $\ket{0}\rightarrow\ket{1}$ rates indicates this effective temperature does not control qubit state populations. Therefore, setting $p_1=0$ and $p_0=1$ captures both the suppression of thermal excitations due to refrigeration and the idiosyncratic asymmetry of QP-induced errors. However, this asymmetry is not a dominant factor in the logical error rate of the surface code. Regardless, this noise channel is applied to single- and two-qubit gates in the form of Kraus operators $K_i$, which in a density-operator formalism act on the density matrix ($\rho$):
$\rho'=\sum_i{K_iU_g \rho U_g^\dagger K_i^\dagger}$,
where $U_g$ is the unitary operator for the selected qubit gate. Gate times and the modeled transmon properties are listed in Table~\ref{tab:transmon_params}.

\begin{figure}
  \subfloat{
    \includegraphics[width=0.95\linewidth]{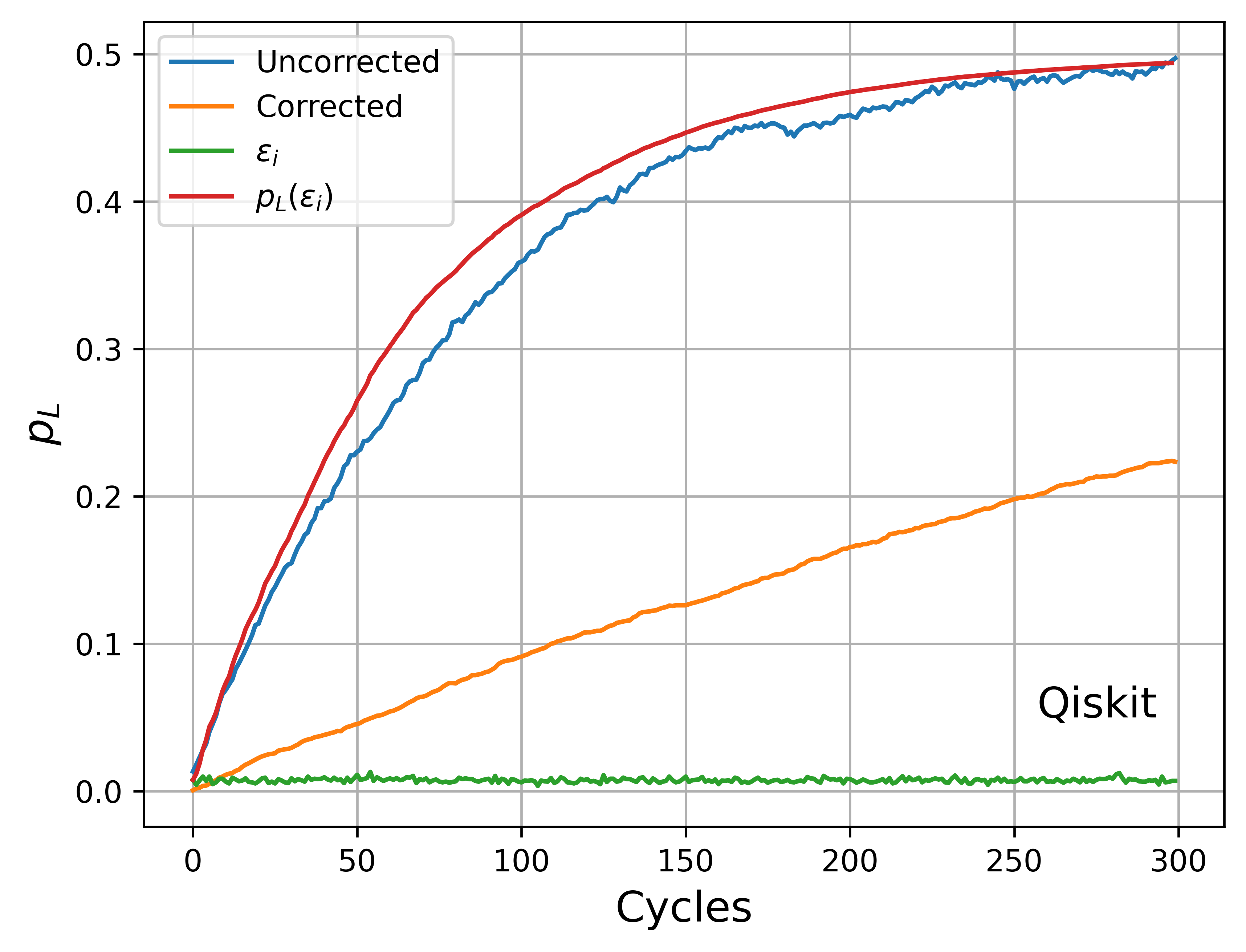}
  }
  \put(-245,160){\bfseries (a)}
  
  \subfloat{
    \includegraphics[width=0.95\linewidth]{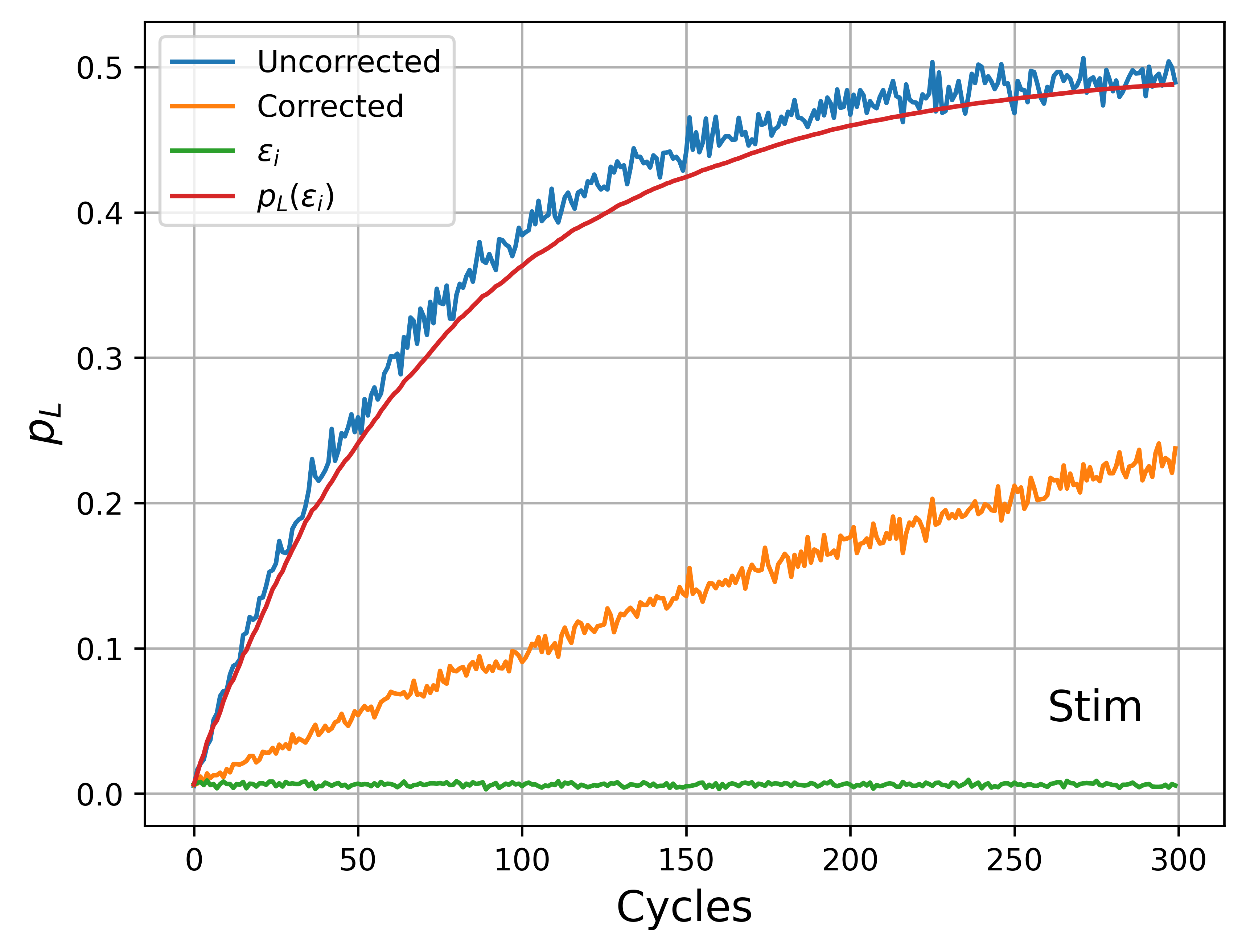}
  }
  \put(-245,160){\bfseries (b)}
  \caption{A comparison of the surface code simulation results using the Python packages (a) Qiskit and (b) Stim. The Qiskit simulation uses the GAD channel of Eq.~\ref{eq:GAD}, while the Stim simulation uses the PTGAD channel for respective noise models. In both cases, $T_1=100$~$\mu$s and $T_2=200$~$\mu$s for each qubit, and the circuit is sampled 4096 times. In the Qiskit case, the syndrome measurements are correlated by passing the state between sequential cycles. For the Uncorrected curves, no error correction is applied, and $p_L$ is measured from the data parity. For corrected curves, the historical syndrome data are decoded by a repetition MWPM decoder (see App.~\ref{app:decoder}), and appropriate bitflips are applied to the $Z$-basis data post measurement. For comparison, the individual cycles are also modeled with single-shot circuit execution, decoding, and correction to calculate the logical error per cycle $\varepsilon_i$. In this latter scenario, states have no dependence on neighboring cycles. The corresponding $p_L(\varepsilon_i)$ can be calculated using Eq.~\ref{eq:pL}. Notably, this method overestimates the error compared to state-chained simulations.\label{fig:PLE}}
\end{figure}

In addition to the Qiskit-based simulation, the same circuit is implemented in Stim \cite{Gidney2021} to achieve a computational speedup. In this case, the noise channel is simplified to a Pauli twirled generalized amplitude damping (PTGAD) channel. Unlike the asymmetric GAD channel, the PTGAD channel has no dependence on $p_0$ or $p_1$ and is therefore rate-symmetric. The Pauli coefficients are expressed as:
\begin{align}
P_X &= P_Y = \frac{1}{4}\bigl(1 - e^{-t_g/T_1}\bigr) \\
P_Z &= \frac{1}{2}\bigl(1 - e^{-t_g/T_2}\bigr) - \frac{1}{4}\bigl(1 - e^{-t_g/T_1}\bigr) \\
P_I &= 1 - P_X - P_Y - P_Z.
\end{align}
Figure~\ref{fig:PLE} compares the Qiskit- and Stim-based simulations. To adequately compare both simulation methods, the readout bits are decoded and corrected using an identical, PyMatching-built minimum weight perfect matching (MWPM) decoder \cite{Higgott2022}. Details about the decoder construction are provided in Appendix~\ref{app:decoder}. Results for the two different simulation platforms are similar. However, Stim-based simulations lead to a statistically insignificant increase to the baseline logical error rates and noisier $p_L$ curves. As both Qiskit and Stim simulations produce nearly identical results, unless otherwise noted, we proceed with Stim-based simulations using the symmetric noise model. 

To quantify the logical error rates, a series of stabilizer circuit rounds is executed using the cycle-evolving noise model. To start, a singular stabilizer cycle with syndrome measurements is first executed without the noise model, resulting in a clean quiescent state \cite{Cleland2022} with known syndromes. 
From this point, two protocols can be established. In Protocol I, the cycles are chained together, so the measured data can be decoded using the full history of syndromes. In Qiskit, this is accomplished by saving the state vector before data measurement then initializing this state vector at the beginning of the next cycle on the updated artificial backend. In Stim, Protocol I is built by appending detector rounds together each with their own noise model. However, measuring the full logical error rate evolution of $N_c$ cycles using this method requires the simulation of $N_c(N_c+1)/2$ cycles. Therefore, in both cases, Protocol I leads to computationally more demanding simulations.
On the other hand, $p_L$ at the $n$th cycle can also be calculated \cite{Acharya2025} from the error rate per cycle $\varepsilon_i$ accumulated in the preceding cycles using the relationship 
\begin{equation}\label{eq:pL}
    p_L = \frac{1}{2}(1-\prod_i^n(1-2\varepsilon_i)).
\end{equation} 
Protocol II employs this relationship to model $p_L$ using error rates $\varepsilon_i$ measured from individual, unlinked stabilizer cycles where the state is self contained within the cycle and unassociated to the states in neighboring cycles. In this case, the noise-free quiescent state with known syndromes is initialized, the noisy circuit is performed followed by a second noiseless syndrome measurement cycle, and the measured data are corrected to deterimine $\varepsilon_i$. This method saves time by simulating only $N_c$ cycles. However, as shown in Figure~\ref{fig:PLE}, Protocol II produces curves matching more closely to the uncorrected state evolution. This behavior is attributed to the data measurement error which occurs after syndrome collection and becomes a dominant factor when the number of cycles is small. Hence, to adequately capture the full QEC effect including decoder and correction, we proceed using Protocol I despite it leading to longer simulation times.

\section{Results}
To briefly summarize the framework for the model detailed in the previous sections, the surface code response to a radiation impact is implemented using two models. The first model, built principally using Geant4 and G4CMP, simulates the generation of QPs following a \muon or $\gamma$-ray strike. The dynamical evolutions to the $T_1$ and $T_2$ times due to phonon propagation are calculated for each qubit with a time resolution $\Delta t=1$~$\mu$s. Each qubit is provided a baseline $T_1^b=100$~$\mu$s and $T_2^b=200$~$\mu$s (i.e.,~$\Gamma_\phi$=0), which is modified per the descriptions in Appendix~\ref{app:T1T2}. These $T_1$ and $T_2$ times are used to calculate the PTGAD noise channel for each gate operation within the surface code operation cycles with duration $t_c=\Delta t$. Because we focus on the effect of radiation impacts, other sources of noise (e.g.,~electrical or two-level systems (TLS)) are omitted. The summary of transmon and gate times is listed in Table~\ref{tab:transmon_params}. The surface code is then modeled in Stim, tracking the syndrome history across sequential cycles with a data measurement in the $Z$-basis in the final cycle. This syndrome history is decoded using the repetition MWPM decoder built from a matching matrix (see App.~\ref{app:decoder}). Once corrections are applied to the data, the data parity is used to measure the logical state, and statistics are collected to measure the evolution of the logical error probability following a radiation impact.

\begin{figure}
  \centering

  \subfloat{
    \centering
    \begin{overpic}[width=0.95\linewidth]{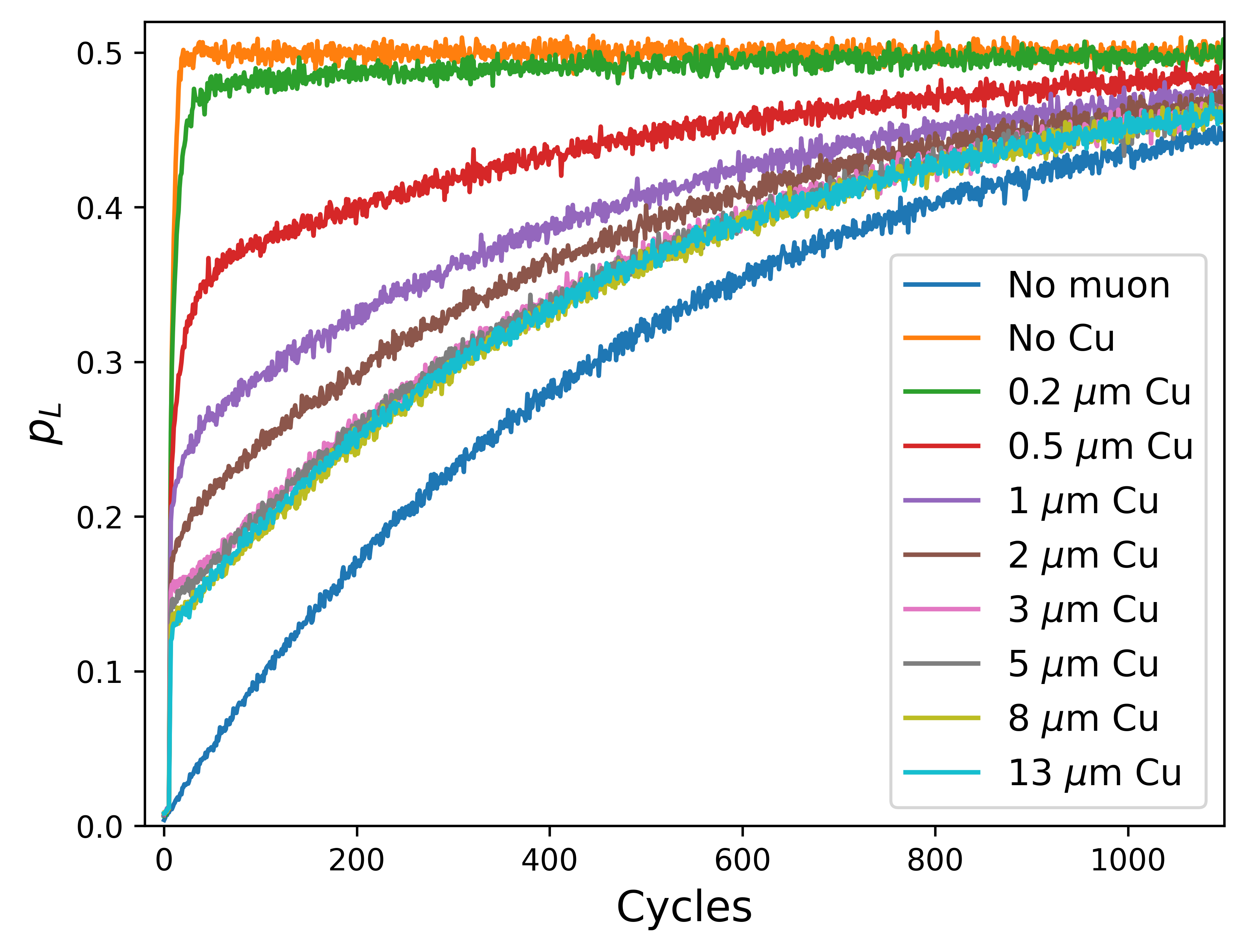}
      \put(0,70){\bfseries (a)}  
    \end{overpic}
  }

  \subfloat{
    \centering
    \begin{overpic}[width=0.95\linewidth]{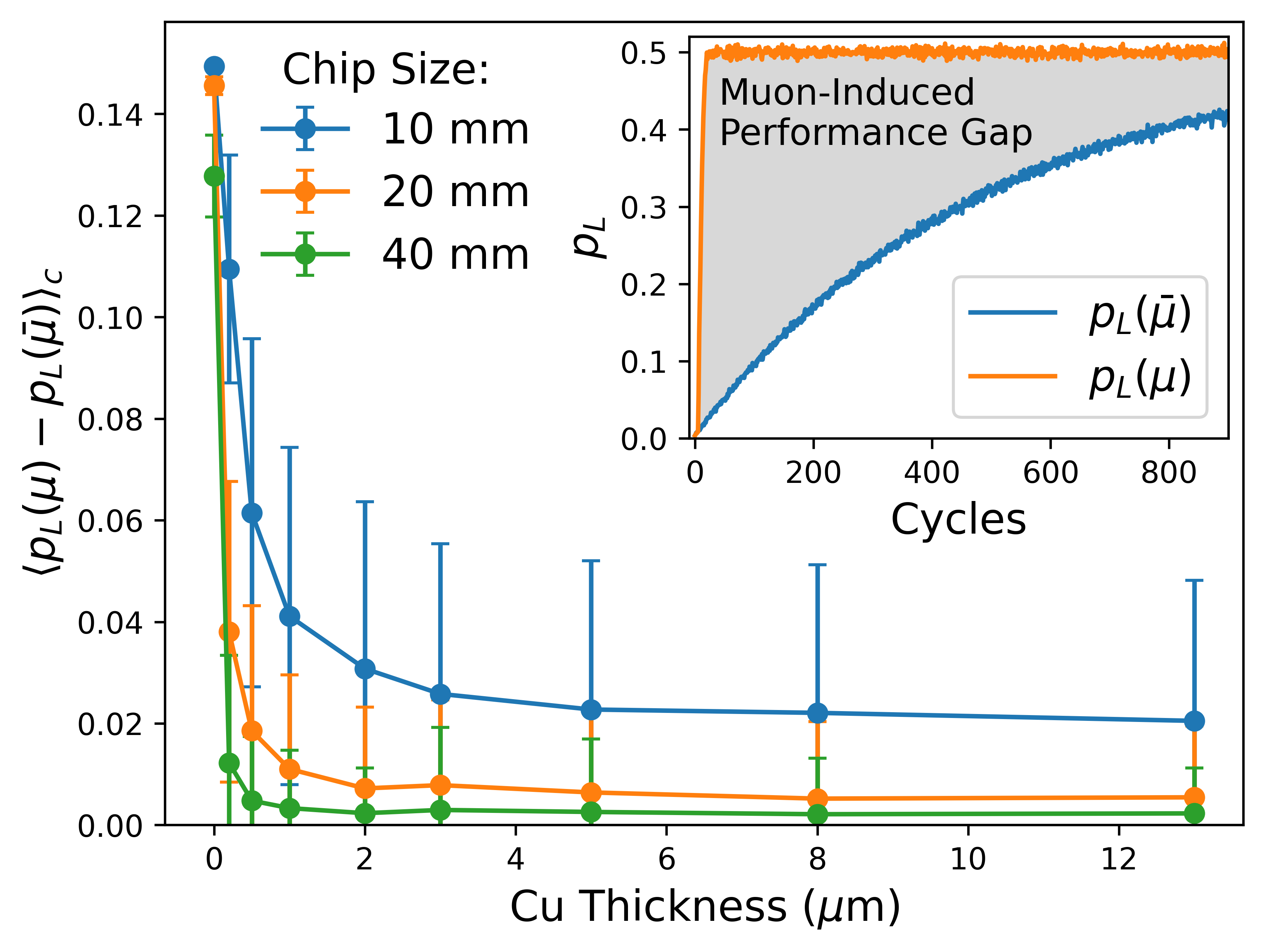}
      \put(0,70){\bfseries (b)}
    \end{overpic}
  }

  \caption{(a) The [[9,1,3]] surface code logical error probability $p_L$ following a muon strike. The simulation consists of the $10\times10$-mm$^2$ chip configuration shown in Figure~\ref{fig:circuit}(a) with corresponding muon strike location. Average $T_1$ and $T_2$ values are calculated from a 100 muon ensemble as described in Appendix~\ref{app:cufigs}. To test the efficacy of phonon downconversion, the chip is modeled with Cu of varying thicknesses covering the underside, such as in \onlinecite{Iaia2022} and \onlinecite{Yelton2024}. (b) The performance gap as a function of Cu thickness and qubit spacing. To sweep qubit spacing, the diagram of Figure~\ref{fig:circuit}(a) is scaled by either a factor of 2 or 4. The chip size is increased to $20\times20$~mm$^2$ and $40\times40$~mm$^2$, while the qubit nearest neighbor spacing is increased from 1~mm to 2~mm and 4~mm. The performance gap $\zeta_c$ is measured over an average of 64 muon strikes normal to the chip surface. (inset) Demonstration of how the performance gap is calculated (shaded region). The difference between $p_L(\mu)$ and $p_L(\bar{\mu})$ is integrated then normalized by the total number of simulated cycles ($N_c\approx1500$). 
  }
  \label{fig:Cu_dep}
\end{figure}

Figure~\ref{fig:Cu_dep}(a) shows some example results of the model. In these examples, the muon strike occurs at the location and chip design shown in Figure~\ref{fig:circuit}(a). The evolution is modeled as if the muon strikes after one noisy cycle, and $N_c^{max}=1500$ total cycles are modeled. The baseline logical error probability calculated without the muon ($p_L(\bar{\mu})$) has a qualitatively similar evolution to those observed in experimental implementations \cite{Acharya2025}, although with a much slower error rate due to the ideal circumstances built into the model. In contrast, the logical error probability with the muon ($p_L(\mu)$) demonstrates a near immediate jump to a 50\% logical error rate, indicating a complete loss of logical state control. This behavior can be attributed to the widespread propagation of phonons across the chip \cite{Yelton2024}.

In addition to the mitigation-free chip design, we can also model the effects of potential mitigation strategies on the surface code operation. As in \onlinecite{Yelton2024}, devices can be modeled with Cu on the backside of the Si die. In this case, Cu rapidly downconverts phonons \cite{Iaia2022} to energies below $2\Delta_{Al}$, the minimum energy required to generate QPs in the qubits' Josephson junctions. Figure~\ref{fig:Cu_dep}(a) also shows the effect of Cu downconvertion on surface code performance following identical muon strikes. As metal thickness increases, the jump size at the muon strike time decreases. Interestingly, unlike the case without mitigation, a brief recovery period is observed when phonon downconversion is present, and $p_L(\mu)$ decreases down to a dip before gradually increasing again. We attribute this behavior to two related effects. First, the effect of phonon downconversion limits the generation of QPs on qubits far from the strike location, and, second, the recovery of the logical state due to correction as syndrome reliability recovers. While qubits have a large burst in error rates, as downconversion proceeds, syndrome measurements become reliable enough to relay \textit{some} information to the decoder. As syndromes are repeated, this information improves the decoder's ability to correct some of the initial burst errors. Hence, the logical error rate reaches a local minimum before the expected low-frequency signal degradation becomes dominant again. As Cu thickness increases and phonon downconversion becomes more efficient, the recovery dip deepens as more reliable syndrome information is messaged to the decoder within each cycle.

The behavior of $p_L(\mu)$ (shown in Fig.~\ref{fig:Cu_dep}(a)) leads to a natural metric to measure the efficacy of both QEC codes and mitigation strategies in either correcting or preventing radiation-induced errors. In this case, we define a performance gap
\begin{equation}
    \zeta_c=\langle p_L(\mu)-p_L(\bar{\mu})\rangle_c,
\end{equation}
which is the cycle average of the difference in logical error rate between radiated and non-radiated test cases. This metric is general and can be applied regardless of the QEC code structure (i.e., arbitrary code distance, homotopy, or decoder), source of radiation ($\gamma$, $\mu$, $\alpha$, etc.), or physical mitigation strategy (downconversion, gap engineering, Si trenching). The Figure~\ref{fig:Cu_dep}(b) inset shows how this metric is defined. Figure~\ref{fig:Cu_dep}(b) mainly demonstrates how $\zeta_c$ can be used to measure the efficacy of mitigation strategies and chip design. Here, Cu thickness and qubit-qubit separation are tuned. In the case of qubit-qubit separation, the qubit arrangement is identical to that shown in Figure~\ref{fig:circuit}(a) but scaled from 1~mm nearest neighbor distance to 2~mm and 4~mm. The chip width is scaled proportionally to accommodate the circuit's increased footprint over the Si surface. For each data point, the average $\zeta_c$ is calculated for 64 muon strikes through the device surface distributed by a two-dimensional Sobol sequence. While chips without any phonon downcoversion show similar $\zeta_c$ regardless of qubit separation, the introduction of Cu rapidly reduces $\zeta_c$. However, qubit-qubit distance must be increased to fully recover the non-radiated performance.

\section{Discussion}\label{sec:Discussion}
The addition of a physical device model to the surface code simulation marks a distinct advantage over more phenomenological models of quasiparticle poisoning. This model provides both realistic spatial and temporal correlations in qubit errors following a radiation event. Additionally, this physical model allows for the consideration of different architecture designs or material platforms and orientations. While not the main focus of this work, Appendix~\ref{app:caustics} explores some of the anomalous behaviors that may arise from phonon caustics. Notably, edge effects, adjustments to phonon length-scales, and anisotropic error distributions may occur which can modify architecture and QEC performance depending on the material. While the immediate effect of phonon caustics is unobserved in our tests due to the statistical variance innate to the Monte Carlo framework, in strongly anisotropic materials (e.g., sapphire), long-range phonon behavior and qubit error correlation can be affected. The physical model established both here and in \onlinecite{Kelsey2023,Yelton2024,Hernandez2025} will likely prove essential for novel architecture design and optimization with respect to radiation-induced errors.

Although our model captures the long-range component of correlated errors from QP generation, there are several avenues for model improvement. Additional physical models for TLS restructuring \cite{Wilen2021,Thorbeck2023,Harrington2025} may be necessary to fully capture the correlated error dynamics closest to the injection site. Furthermore, per Section~\ref{sec:QECModel}, the thermal populations are set as $p_0=1$ and $p_1=0$ in the GAD noise channel. This better models the QP dynamics by reflecting both the thermal statistics of a sufficiently cold device, as well as the asymmetry of QP-induced error rates. As explained in \cite{McEwen2022}, this rate asymmetry occurs due to the fast cooling of QPs to the superconducting gap edge so that they can only absorb energy from the qubit. Under this explanation, the noise model may need adjustment in the case of constant phonon injection \cite{Iaia2022} or steady-state phonon generation from the refrigerator pulse-tube \cite{Yelton2025}. In these scenarios, the continuous generation of hot QPs may modify this rate asymmetry, leading to increased occupation of excited states more in line with Maxwell-Boltzmann statistics \cite{Jin2015} as measured by $T_{eff}(x_{qp})$. Given the elevated $T_{eff}(x_{qp})\sim250$~mK observed in our simulations, these scenarios may also lead to noise leakage with occupancy of higher energy states ($\geq\ket{2}$), which is not captured in our model. Additional experiments examining rate symmetry with steady-state phonon injection may close this knowledge gap. Furthermore, while the asymmetry may not affect overall error rates of the surface code, this analysis may change in the context of noise leakage. Recent work \cite{Kurilovich2025} has noted a brief but measureable time shortly after QP-generation when QPs have energies above the gap edge. This leads to a brief symmetry in error rates, but might also lead to a brief increase in noise leakage.

Simulations to test the efficacy of phonon downconversion as a mitigation strategy for radiation-induced errors demonstrate that additional limitations on qubit-packing density are required to fully suppress errors. While this method reduces error rates substantially (i.e.,~$\zeta_c\to0$) compared to devices with no mitigation strategy, qubit separation distance must be increased to fully suppress the effects of correlated errors. Furthermore, the majority of the benefit occurs with only a thin layer of Cu, while increasing the thickness of the downconversion metal provides only diminishing returns in performance. In other words, increasing the downconversion metal thickness does not allow for tighter packing of qubits on the chip. However, this also means that a thin layer ($\lesssim 1$~$\mu$m) of a normal metal or low-gap superconductor \cite{Henriques2019} is likely sufficient to gain most of the benefit from this strategy. For example, in the 40-mm chip design, 500~nm (200~nm) of Cu is expected to be only 2$\%$ (8$\%$) less effective than 13~$\mu$m of Cu and well within the uncertainty range for the sampled muon ensemble (see App.~\ref{app:cufigs} for details). In this light, the implementation of phonon downconverting metal films is still compatible with the standard CMOS (Complementary Metal Oxide Semiconductor) fabrication techniques as the electroplating method used in \onlinecite{Iaia2022} to create thick films of Cu may be unnecessary. Instead, such films can be sufficiently thin that Ar-plasma sputtering alone can be used. Furthermore, as higher-distance codes are generally more resilient to error, the optimal qubit-packing density to withstand radiation-induced errors may itself increase with the code distance. A more detailed study with experimental validation may be necessary to determine this dependence.

\section{Conclusion}
We have established a method for modeling QEC performance in the presence of radiation impacts on a superconducting quantum processor. Although only a simple 17-qubit [[9,1,3]] surface code is demonstrated, this method is generalizable to arbitrary architectures or code structures. Additionally, we have confirmed that a Pauli-twirled GAD quantum channel can adequately model the error rates during a radiation impact despite the differences in error rate symmetry between GAD and PTGAD channels. Lastly, we have introduced a new metric, $\zeta_c$, to measure the performance of QEC codes and physical error mitigation strategies for radiation-induced errors. The usefulness of $\zeta_c$ has been demonstrated for the mitigation test case of phonon downconversion using flipside Cu. In this regard, machine-learning–driven optimization and active learning methods applied toward reducing $\zeta_c$ can be used to efficiently search the enormous design space of quantum processors \cite{alexander2023accelerating}. By leveraging uncertainty-aware surrogate models \cite{qian2023knowledge}, this approach may actively steer the expensive physical simulations established in this manuscript toward the most informative design candidates and enable the discovery of architectures with improved resilience to radiation-induced errors.

\begin{acknowledgments}
This work was supported by the Laboratory for Physical Sciences through Strategic Partnership Project No.~EAOC0206928. Brookhaven National Laboratory is supported by the U.S. Department of Energy's Office of Science under Contract No. DE-SC0012704. This work was also supported by resources provided by the Scientific Computing and Data Facilities (SCDF), a component of the Computing and Data Sciences Directorate at Brookhaven National Laboratory. We also thank C.~Plata for editorial review of the manuscript. AI Disclaimer: Both ChatGPT and Claude Sonnet were used to suggest code modifications to human-written codes for the purposes of troubleshooting and error handling. Adopted code changes were first validated by the authors. The manuscript text, figures, and contents were written and generated solely by the authors themselves.
\end{acknowledgments}

\appendix

\section{Quasiparticle Model}\label{app:QPM}
As discussed in the main text, we adapt the recent framework developed by Yelton et al. \cite{Yelton2024} to model the quasiparticle (QP) density ($n_{qp}$) evolution after a \muon or $\gamma$ ray impact. This model is based on a Geant4 and G4CMP Monte Carlo simulation, whose output is used to evaluate the generation term $g_{qp}(t)$ for QPs in the ordinary differential equation (ODE):
\begin{equation}
    \frac{dn_{qp}}{dt}=-rn_{qp}^2-sn_{qp}+g_{qp}(t).
\end{equation}
Here, $r$ is the QP recombination rate, and $s$ is the decay rate. While these rates can vary from qubit to qubit in practical devices, we choose nominal values of $s=0.05$~$\mu$s$^{-1}$ and $r=25\times 10^{-6}$~$\mu$s$^{-2}$ for all qubits, consistent with the range of values found in ~\cite{Yelton2024}.

The device itself is modeled as a square Si die with dimensions $10\times10\times0.525$~mm$^3$. The die's surface is blanketed with a 75-nm thick Nb film. Transmon junctions are modeled as $10\times10\times0.1$~$\mu$m$^3$ Al patches distributed across the die. For the surface code arrangement used in the main text, there is a $\sim1$~mm pitch between qubit junctions. Radiation impacts are modeled as ionization (\muon) or Compton scattering ($\gamma$ ray) processes in the Si for the generation of \eh pairs. These pairs then participate in the Neganov-Trofimov-Luke (NTL) process \cite{Kelsey2023} to generate phonons. This method differs from \cite{Yelton2024}, which directly injects \eh pairs to model $\gamma$ ray impacts. Modeling the full ionization process considers the transience of muons during the ionization process and pair generation. Moreover, the Compton scattering event for $\gamma$ rays also creates recoil electrons \cite{Norcini2022}, which themselves further ionize \eh pairs as they traverse the Si. In such scenarios, a purely stationary generation point for \eh pairs may not fully characterize the dynamics of qubit error generation, especially when mitigation strategies limit phonon propagation and not muon or recoil-electron propagation.

Computationally evaluating the ODE provides the normalized QP density $x_{qp}(t)=n_{qp}(t)/n_{cp}$, where $n_{cp}$ is the density of Cooper pairs. To keep the model internally consistent, we adjust $n_{cp}$ to reflect the modified Al gap energy $\Delta_{Al}=191\pm1$~$\mu$eV used in \cite{Yelton2024}. As defined in \cite{Catelani2011}, $n_{cp}=2g_1(0)\Delta$, where $g_1(0)$, is the single electron density of states at the Fermi surface. Using the free electron gas model from \cite{Kittel2005}, $g_1(E)=\frac{(2m_e^3(E-E_{con}))^{1/2}}{2\pi^2\hbar^3}$, a conduction band energy of $E_{con}=11.3\pm0.4$~eV (averaging the range of values in \cite{Kittel2005,Levinson1983}), and $\Delta_{Al}$, we find $n_{cp}=(4.38\pm0.09)\times10^6$~$\mu$m$^{-3}$, a $\sim10\%$ increase from the values used in \cite{Wang2014,Yelton2024}. This increase stems from the increased $\Delta_{Al}$ value. However, as noted in \cite{Yelton2024}, $\Delta_{Al}$ is thickness dependent and will likely change with design and fabrication details. Therefore, $n_{cp}$ should also be reevaluated depending on the system.

\section{Modeling Amplitude and Dephasing Errors}\label{app:T1T2}
Because the surface code depends on both $X$- and $Z$-parity checks, dephasing error rates due to QP generation need to be considered. To model dephasing rates, we use the formulas provided by \cite{Catelani2012} for transmon qubits. Noting these equations are derived in a quasi-equilibrium approximation with an effective temperature $T_{eff}$ describing the normalized QP density $x_{qp}$, the equilibrium description $x_{qp} = \sqrt{\frac{2\pi k_B T_{eff}}{\Delta_{Al}}}e^{-\Delta_{Al}/k_B T_{eff}}$ \cite{Barends2011} can be used to translate the simulation-derived $x_{qp}$ into $\Gamma_1$ and $\Gamma_\phi$. The inverse solution $T_{eff}(x_{qp})$ can be expressed using the product logarithm, which corresponds to the principle ($k=0$) branch of the Lambert $W_k(\cdot)$ function:
\begin{equation}
  T_{eff} = \frac{2\Delta_{Al}}{k_BW_0(4\pi/x_{qp}^{2})}.
  \label{eq:effectivetemp}
\end{equation}

\begin{figure}[ht]
  \centering
  \includegraphics[width=\linewidth]{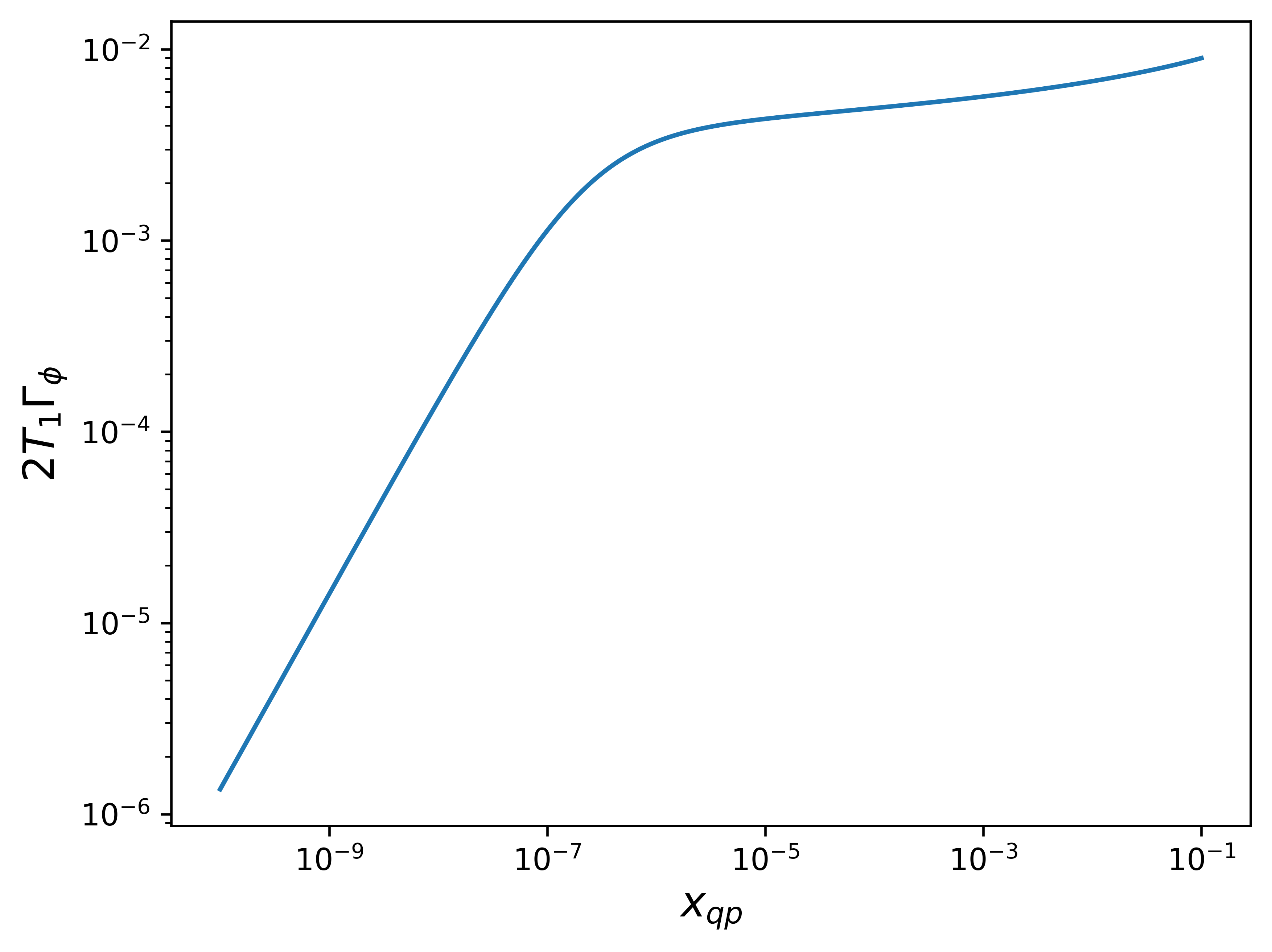}
  \caption{$2T_1\Gamma_\phi$ measures the comparison between the dephasing and relaxation rates as a function of $x_{qp}$. The minimum and maximum range of $x_{qp}$ covers the ranges observed in the simulation data. These values have been calculated using the transmon parameters listed in Table~\ref{tab:transmon_params}.}
  \label{fig:dephasing}
\end{figure}

Changes to the qubit relaxation rate can then be expressed in terms of $x_{qp}$ as
\begin{equation}
\begin{split}
\Delta \Gamma_1
  &= 2 \sqrt{\frac{\omega_{01} k_BT_{eff}}{\pi\hbar}}\,e^{-\Delta_{Al}/k_BT_{eff}} \\[4pt]
  &= 2 \sqrt{\frac{2\omega_{01}\Delta_{Al}}{\pi\hbar W_0\!\bigl(4\pi/x_{qp}^2\bigr)}}\,
  e^{-\tfrac12\,W_0\!\bigl(4\pi/x_{qp}^2\bigr)} \\[4pt]
  &= \frac{x_{qp}}{\pi}\sqrt{2\omega_{01}\Delta_{Al}/\hbar},
\end{split}
\label{eq:Gamma1}
\end{equation}
using the definition of the product logarithm $W_0(z)e^{W_0(z)}=z$ to derive the linear dependence on $x_{qp}$, which is used in \cite{Wang2014, Iaia2022}. This same method can be used to find the changes to the dephasing rate:
\begin{equation}
\begin{split}
\Delta \Gamma_\phi
  &= \frac{E_C k_B T_{eff}}{\pi\hbar\Delta_{Al}}e^{-\Delta_{Al}/k_B T_{eff}} \\[4pt]
  &= \frac{E_C x_{qp}^2}{2 \pi^2\hbar}e^{\tfrac{1}{2}W_0\!\bigl(4\pi/x_{qp}^2\bigr)}.
\end{split}
\label{eq:Gamma_phi}
\end{equation}
Thus, the QP-induced dephasing rate can be expressed in terms of the $x_{qp}$. $T_2$ can then be calculated as $1/T_2=1/2T_1 + \Delta \Gamma_\phi +\Gamma_\phi^b$, where $\Gamma_\phi^b$ is the baseline dephasing rate, which is assumed negligible (i.e., $\Gamma_\phi^b=0$).  As noted in \cite{Catelani2012}, the dephasing rate contribution from QPs is expected to be much smaller than the relaxation rate. This is verified by the numerical evaluation of $2T_1\Gamma_\phi$, shown in Figure~\ref{fig:dephasing}. Nevertheless, for larger $E_C$ values, the rate of dephasing errors may become noticeable at the higher levels of $x_{qp}$ achieved following radiation impact events.

\section{Error Decoding} \label{app:decoder}

To determine the logical error probability, errors are decoded using a Minimum Weight Perfect Matching decoder built with PyMatching \cite{Higgott2022}. Bit errors are corrected after $Z$-basis data measurement. The matching matrix to decode the $ZZZZ$ syndromes is determined by the circuit \cite{deMartiiOlius2024} shown in Figure~\ref{fig:circuit}(b):
\[
H_Z =
\begin{bmatrix}
1 & 0 & 0 & 1 & 0 & 0 & 0 & 0 & 0 \\
0 & 1 & 1 & 0 & 1 & 1 & 0 & 0 & 0 \\
0 & 0 & 0 & 1 & 1 & 0 & 1 & 1 & 0 \\
0 & 0 & 0 & 0 & 0 & 1 & 0 & 0 & 1
\end{bmatrix}.
\]
To avoid notation confusion, the subscript here denotes the parity being checked ($Z$-parity) rather than the type of error being corrected ($X$-error). This matching matrix is then used by PyMatching to construct the syndrome graph necessary to decode errors from syndrome measurements. While each cycle $i$ provides a set of $ZZZZ$ syndrome measurements $S_i^Z$, the decoder graph is constructed with repetition, and the syndrome differences $D_i^Z$ between neighboring cycles
\begin{equation}
    D_i^Z = S_{i - 1}^Z \oplus S_i^Z
\end{equation}
are the detectors for errors and provided to the decoder instead of the raw syndrome data. In this case, the first noisy cycle ($i=0$) is compared to the noise-free syndrome measurement determined when the noise-free quiescent state is first prepared. Like the $Z$-syndromes, the $XXXX$ syndromes $S_{i}^X$ can be decoded in PyMatching using the matching matrix
\[
H_X =
\begin{bmatrix}
0 & 1 & 1 & 0 & 0 & 0 & 0 & 0 & 0 \\
1 & 1 & 0 & 1 & 1 & 0 & 0 & 0 & 0 \\
0 & 0 & 0 & 0 & 1 & 1 & 0 & 1 & 1 \\
0 & 0 & 0 & 0 & 0 & 0 & 1 & 1 & 0
\end{bmatrix}.
\]
Once decoded, the operations can be applied to correct phase errors and data measured in the $X$-basis. Regardless of the type of error, in our study, data bits are corrected after measurement based on the decoded syndromes.

\begin{figure}
  \centering
  \includegraphics[width=\linewidth]{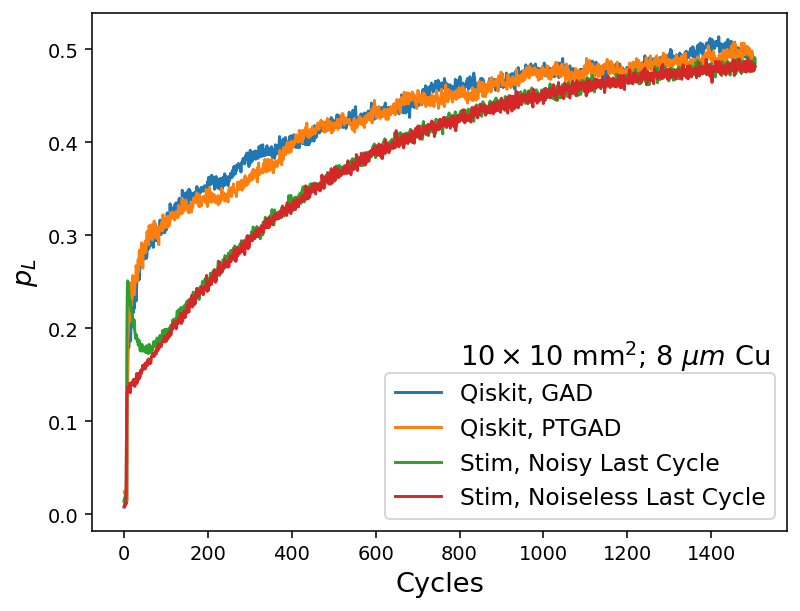}
  \caption{Tests of the surface code implementation for a device with phonon downconversion. $p_L(\mu)$ is measured for both Qiskit and Stim implementations. In the Stim-based simulation, an additional noiseless cycle is added to the end of the circuit to aid the decoder. Without this last noiseless cycle, $p_L(\mu)$ features a pronounced dip. To track the source of this behavior, Qiskit simulations are also implemented with GAD and PTGAD noise models. Both Qiskit curves are implemented with a last noiseless cycle, however, no difference is observed when the last cycle is noisy. Curves with a noisy last cycle are not shown due to the complete overlap with the curves present in the figure. For reference, $p_L({\bar{\mu}})$ is also included to demonstrate that without an error burst, the two simulations agree.}
  \label{fig:decoder_tests}
\end{figure}

To aid in error decoding, an additional noiseless cycle is appended to the end of the surface code for results in the main text. Without this last noiseless cycle, the Stim-based simulations exhibit a pronounced spike and dip feature within the first hundred cycles as shown in Figure~\ref{fig:decoder_tests}. Interestingly, this feature is absent in the Qiskit model, which remains unaffected by the presence of the noiseless round (not shown). To test if this feature is due to the difference in noise models, we rerun the Qiskit simulation using Pauli noise (PTGAD). The PTGAD Qiskit simulation remains consistent to the GAD simulation, and therefore the occurrence of this dip cannot be attributed purely to the noise model.
Regardless of the noise model, Qiskit simulations seem to produce higher $\zeta_c$ values compared to Stim. Interestingly, this difference between Qiskit and Stim is not observed in $p_L(\bar{\mu})$ (Figure~\ref{fig:PLE}), and is only observed in the case of $p_L(\mu)$ where the error rate has a sudden burst. The source of both of these discrepancies is unknown, but could be due to difference in state tracking between the two packages.

\section{Effects of Caustics and Anisotropy}
\label{app:caustics}

As mentioned in the main text, our model differs from comparable studies through the use of a quantitative physical model of the device under radiation duress. While previous studies have used qualitative models for errors based distance from the impact center, these models do not consider the role of phonon caustics \cite{Hernandez2025}, anisotropy, or edge effects. Here we test the role of all three on the distribution of phonons across a device.

\begin{figure}
  \centering
  \subfloat{%
    \centering
    \begin{overpic}[width=0.85\linewidth]{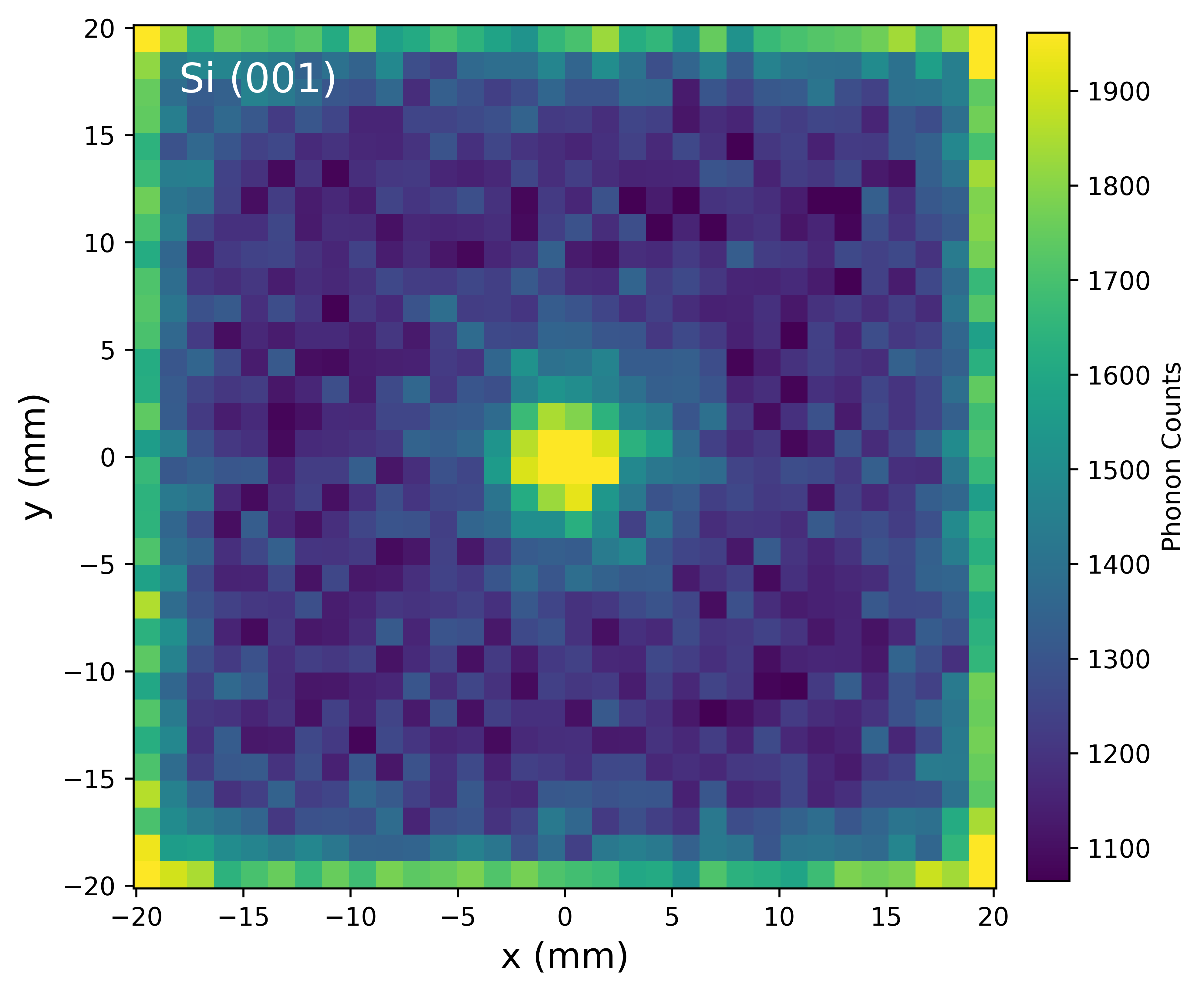}
      \put(-1,75){\bfseries (a)}
    \end{overpic}
  }\vspace{-3mm}
  \subfloat{%
    \centering
    \begin{overpic}[width=0.85\linewidth]{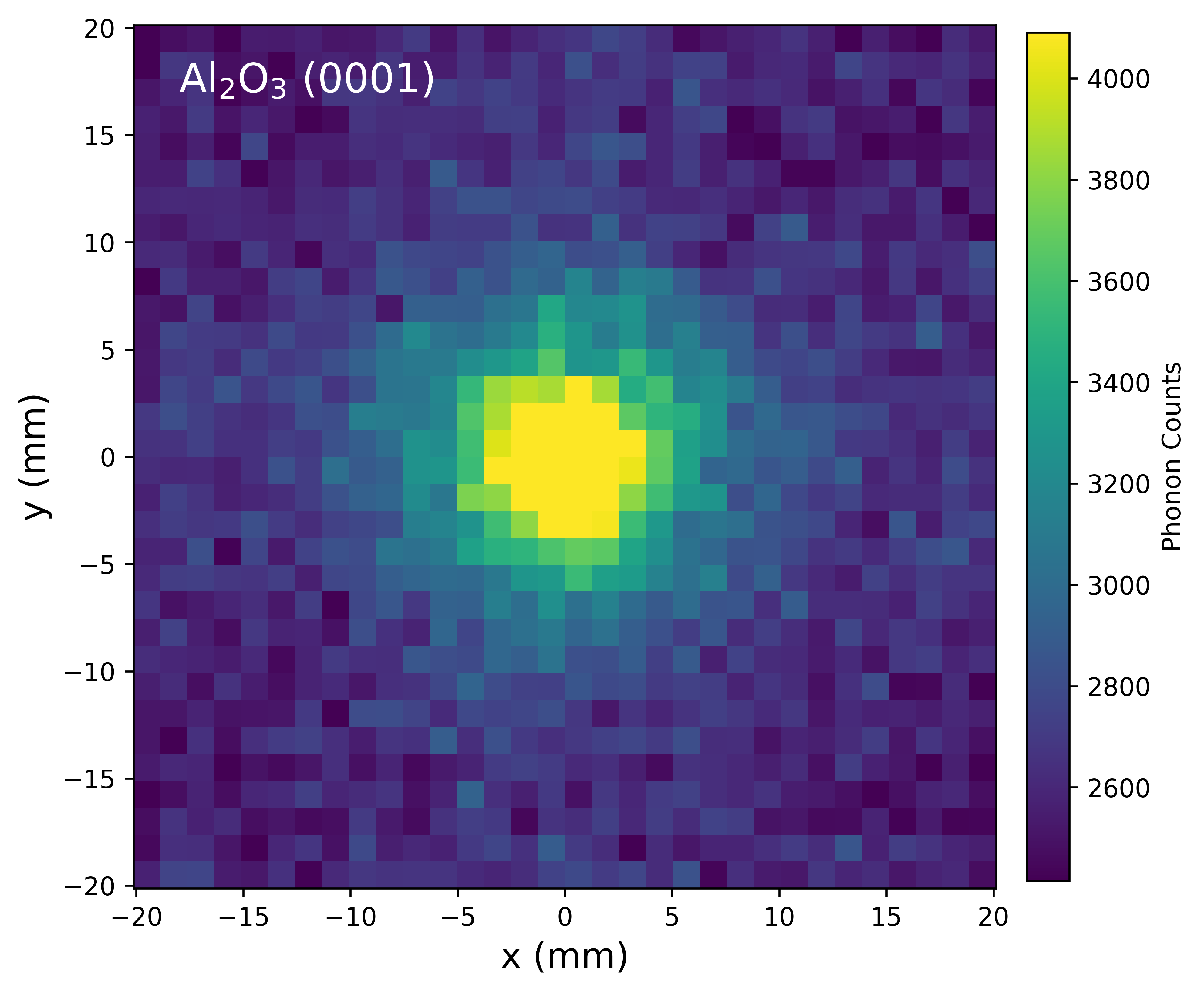}
      \put(-1,75){\bfseries (b)}
    \end{overpic}
  }\vspace{-3mm}
  \subfloat{%
    \centering
    \begin{overpic}[width=0.85\linewidth]{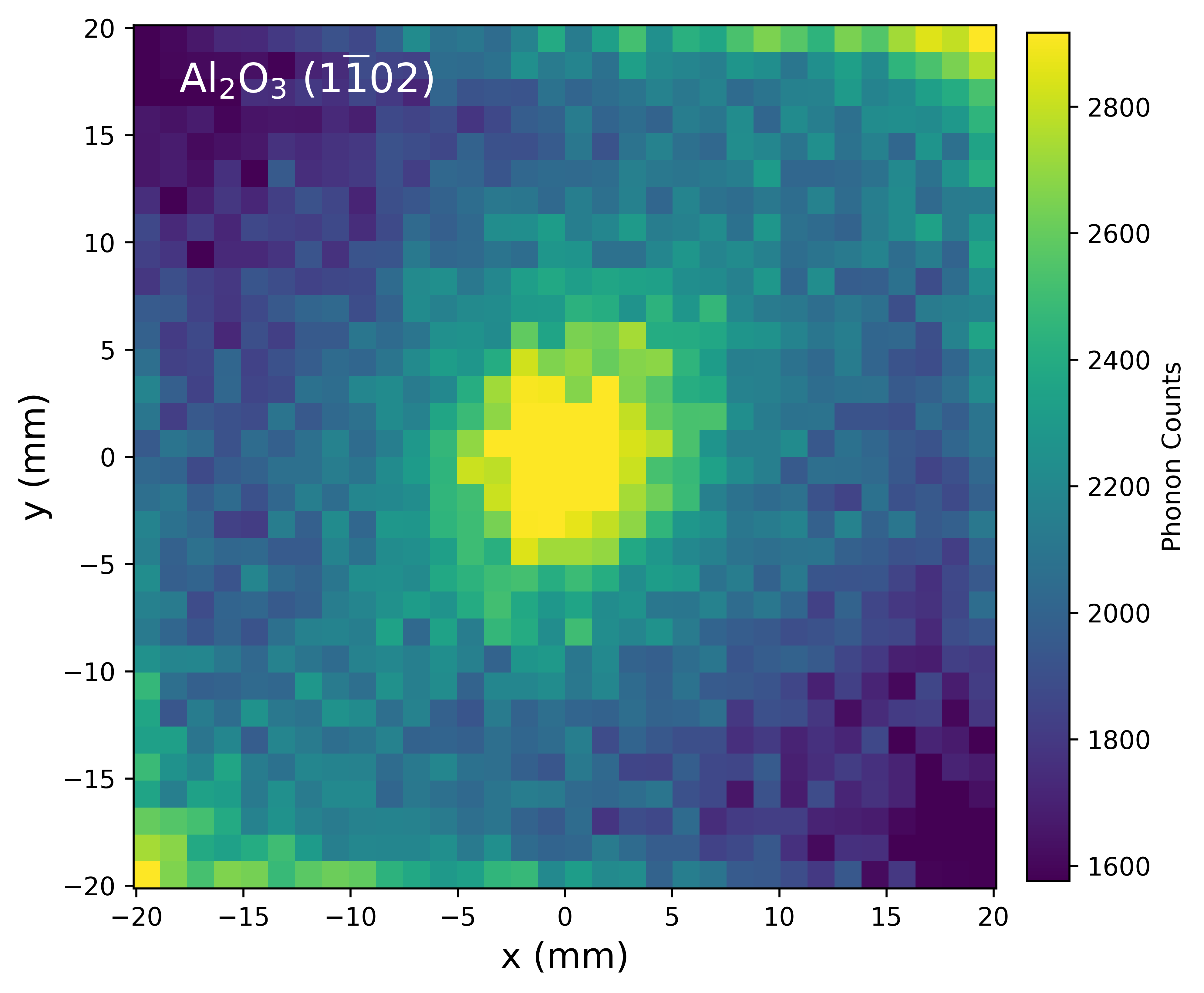}
      \put(-1,75){\bfseries (c)}
    \end{overpic}
  }\vspace{-3mm}
  \caption{(a) The distribution of phonons in a (001)-oriented Si chip. The device is modeled with a $32\times32$ grid of Al detector pads distributed across a $40\times40$ mm$^2$ chip. Like the simulations in the main text, a Nb film covers the remaining surface area. 10M $10$~meV phonons are injected into the center of the chip, and the total number of phonons which strike each Al pad are counted.
  (b) The distribution of phonon strikes for (0001)-oriented, C-plane sapphire.
  (c) The distribution of phonon strikes for (1$\bar{1}$02)-oriented, R-plane sapphire. The anisotropy creates a noticeable asymmetry in the number of phonons detected in the corners of the device. For (a)-(c) the color bar ranges are reduced to enhance the contrast.}
  \label{fig:caustics}
\end{figure}

To test how phonons are distributed throughout the device, we model a $40\times 40$~mm$^2$ Si chip with a $32\times 32$ grid of Al junction pads distributed across the surface. As in the main text simulations, the Si has (001), C-plane orientation, and a Nb film covers the top-side remaining surface area. The underside of the chip is bare without a downconversion material. To test the distribution of phonons, 10M 10~meV phonons are generated in the center of the device. This energy corresponds to the maximum phonon energy observed in the muon strike simulations. These phonons rapidly downconvert through both intrinsic decay and interactions with the Nb film, generating a distribution of lower-energy phonons, which propagate across the device. Figure \ref{fig:caustics}(a) shows the distribution of phonon counts, where phonons are absorbed by the grid of Al detectors.

Two major characteristics are observed. First, there is a large peak in phonon counts below the injection site in the center of the chip. This behavior is expected, as phonons are most likely to strike the junction pads closest to the injection site. Further away from the injection site, the phonon count decreases to $\sim30\%$ of the maximum value near the center. Here, we note that the maximum phonon count is $\sim3$k just under the injection site, however the colorbar range in Figure \ref{fig:caustics}(a) is cut off to enhance visual contrast away from the center. Regardless, the second major feature is observed along the edge of the device. Here, the phonon count once again \textit{increases} to $\sim50\%$ of the maximum value. This can most likely be explained by reflections of the phonons at the boundary of the device, though there may be additional caveats as explained below. Comparable to the simulations in \onlinecite{Yelton2024}, the boundary is modeled with a 0.02 absorption coefficient with completely diffuse scattering.

Less notable in Figure \ref{fig:caustics}(a) are indications of phonon caustics. While these caustics undoubtedly occur within the simulation---we have independently confirmed the observations of \onlinecite{Hernandez2025} (not shown)---they are likely masked by the statistical noise and junction separation. In this case, the effects of the caustics are not strong enough to be observed in Si. However, the effects of caustics may be stronger in systems with large anisotropies. To test this hypothesis, we perform the same simulations in sapphire (Al$_2$O$_3$) with different orientations. Aside from swapping the substrate material, the simulations remain otherwise unchanged from the Si case.

Figure \ref{fig:caustics}(b) shows the phonon counts for a sapphire simulation with (0001), C-plane orientation. The simulation results differ from Si through two observations. Firstly, the total number of phonon counts is much larger and the bright spot at the center is broader. From the material files in G4CMP \cite{Kelsey2023,Hernandez2025}, the phonon lifetimes can be calculated from the scattering and decay rates ($\Gamma_s$ and $\Gamma_d$, respectively) as $\tau_{ph}=1/(\Gamma_d(\omega) + \Gamma_s(\omega))$. At $\omega = 2\Delta_{Nb}/\hbar$, $\tau_{ph}$ is approximately three times larger in sapphire than Si. The increased lifetimes mean that phonons have more time to interact with the junctions. Combined with the increased velocity, phonons in sapphire also travel further before decay. Secondly, the enhanced phonon count near the edge of the device is absent. Therefore, this edge effect may have additional symmetry constraints even with diffuse scattering. Alternatively, the effect may be generally weak and therefore masked by the increased number of phonon counts. Regardless, the C-plane sapphire demonstrates a radially symmetric phonon count profile across the device as observed in Si. 

To test the effects of anisotropy, a different plane of sapphire, such as the M-, A-, or R-planes, can be selected. Figure \ref{fig:caustics}(c) shows the distribution of phonon counts for the (1$\bar{1}$02), R-plane orientation. While the features exhibited in C-plane sapphire are still generally present, a noticeably asymmetry is observed in the chip corners. Indeed, the upper left and lower right corners have only $\sim 50\%$ of the phonon counts as the lower left and upper right corners. This behavior is also observed in the M- and A-planes (not shown) and arises from the asymmetry in the phonon caustics. Therefore radially symmetric error models may not be generalizable for all sapphire-based devices or other systems with strong caustic anisotropies. Exploiting these types of asymmetries may lead to novel architectures that better resist the effects of radiation, especially in materials with reduced $\tau_{ph}$.

\section{Analysis of Downconversion Efficacy} \label{app:cufigs}
\begin{figure}
  \centering

  \subfloat{
    \centering
    \begin{overpic}[width=0.95\linewidth]{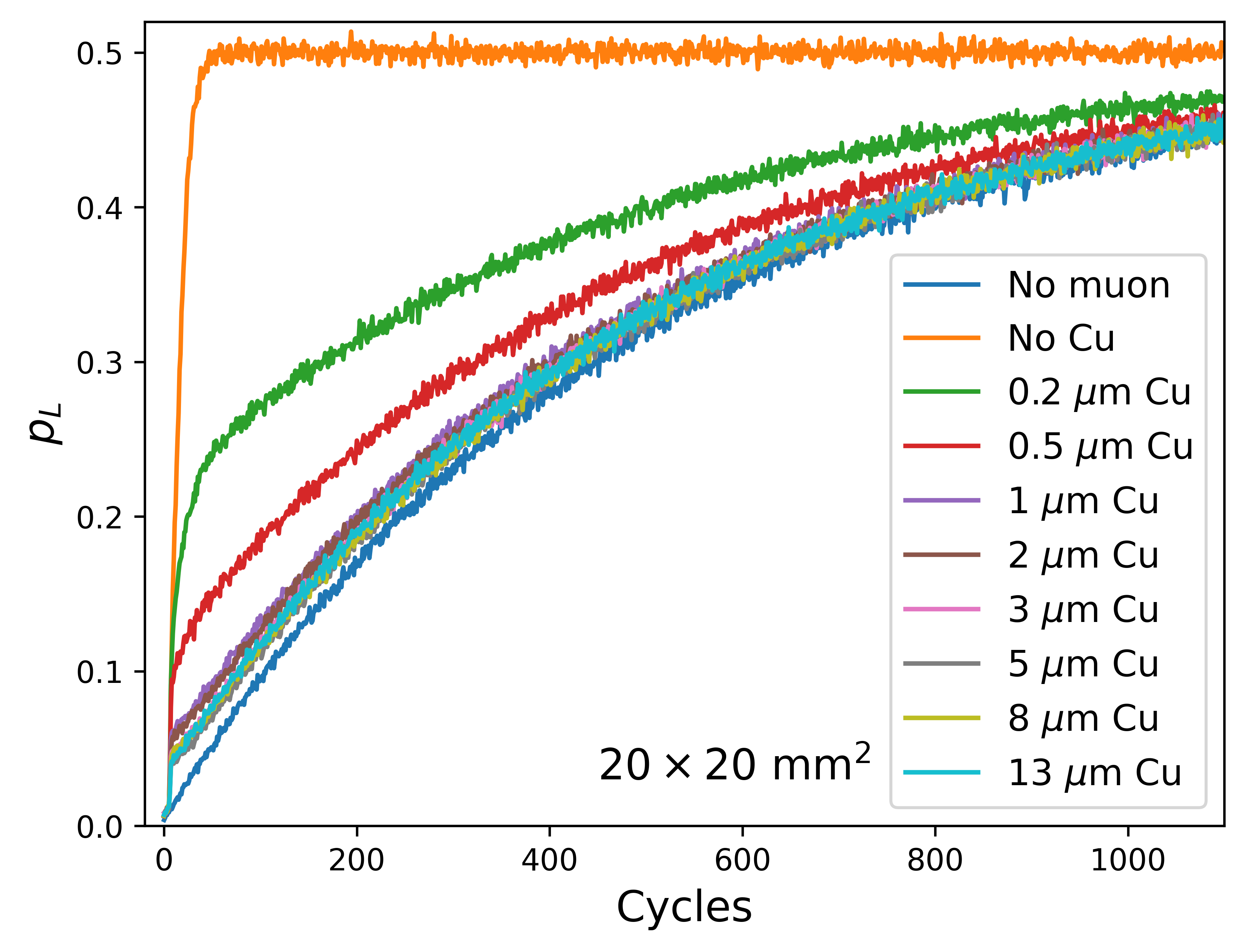}
      \put(-1,70){\bfseries (a)}  
    \end{overpic}
  }

  \subfloat{
    \centering
    \begin{overpic}[width=0.95\linewidth]{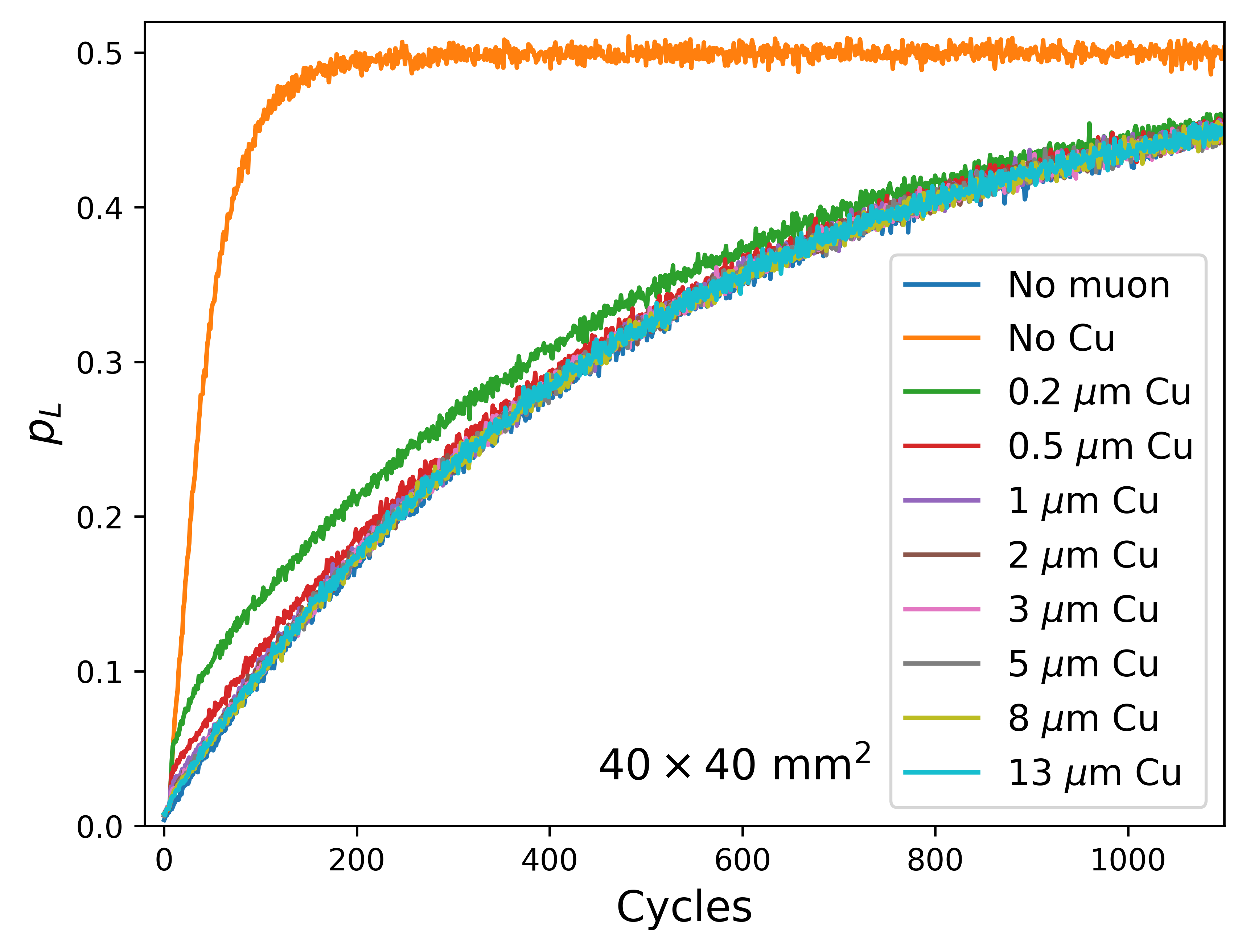}
      \put(-1,70){\bfseries (b)}
    \end{overpic}
  }

  \caption{Sample $p_L$ curves for the (a) $20\times20$~mm$^2$ and (b) $40\times 40$~mm$^2$ chips. For both devices, the qubits are arranged identically to Figure~\ref{fig:circuit}(a) inset but are scaled by a respective factor of two or four. Likewise, the muon strike location is also scaled from (1.7, 1.7) mm in Figure~\ref{fig:circuit}(a) inset to (3.4, 3.4)~mm and (6.4, 6.4)~mm. For both cases, as Cu metal thickness increases, the $p_L(\mu)$curves start to approach the $p_L(\bar{\mu})$ curve, and $\zeta_c$ approaches zero.}
  \label{fig:Cu_dep2}
\end{figure}

To generate the data in Figures~\ref{fig:Cu_dep}(a) and \ref{fig:Cu_dep}(b), we use two different protocols. In Figure~\ref{fig:Cu_dep}(a), the data show the $p_L(\mu)$ for one particular \muon trajectory given varying degrees of Cu thickness. However, to eliminate statistical variations in energy deposition, $T_1$ and $T_2$ are calculated from the average of 100 strikes at the noted location in the Figure~\ref{fig:circuit}(a) inset. This is done within the Geant4/G4CMP simulation by setting \texttt{\detokenize{/g4cmp/producePhonons}} and \texttt{\detokenize{/g4cmp/producePhonons}} each to 0.01 and \texttt{\detokenize{/run/beamOn}} to 100 \cite{Kelsey2023}. The resulting output file then contains the phonon data for the collective 100 \muon strikes, which each individually contribute $\sim1\%$ of the phonons. These phonon data then are used to calculate $g_{qp}(t)$ as described in \onlinecite{Yelton2024}. $p_L$ is calculated from the Stim-based simulations with 16,384 runs of the circuit featuring these $T_1$ and $T_2$ values. Figures~\ref{fig:Cu_dep2}(a) and \ref{fig:Cu_dep2}(b) respectively depict the $p_L$ for the $20\times20$~mm$^2$ and $40\times 40$~mm$^2$ devices. In these cases, the geometry -- including the muon strike location -- is scaled proportionally by a respective factor of two or four.

For Figure~\ref{fig:Cu_dep}(b), $\zeta_c$ is instead averaged for an ensemble of 64 muon strikes wherein the strike location is varied. In this case, the strike locations are generated by a two-dimensional Sobol sequence that is scaled by the chip dimensions. For each muon, phonon statistics are fully sampled, and $\zeta_c$ is calculated. The $\zeta_c$ values are then averaged for the 64 \muon ensemble, and the error bars represent the standard deviations. From these data, we conclude that a 500-nm thick film of Cu is only $2\%$ less effective than a 13-$\mu$m thick film of Cu for the 40-mm wide chip. This estimation is calculated by comparing the mitigation efficacy of the film downconversion relative to the device without the Cu film. The relative mitigation efficacy is calculated as
\begin{equation}
    \delta\zeta_c(t_{Cu}) = [\zeta_c(t_{Cu})-\zeta_c(0)]/\zeta_c(0),
\end{equation}
where $t_{Cu}$ is the Cu film thickness. Using this formulation, $\delta\zeta_c(500~\mathrm{nm})\approx0.96$, and $\delta\zeta_c(13~\mu\mathrm{m})\approx0.98$. Therefore, we conclude the efficiency of the downconversion effect only differs by $2\%$ between a 500-nm film and a 13-$\mu$m film. For a slightly thinner $\delta\zeta_c(200~\mathrm{nm})\approx0.90$, the difference increases to 8$\%$, which is still nominal compared to the standard deviation of $\zeta_c$ values calculated from the ensemble of \muon strikes. Based on this analysis, we conclude that thin films made of Cu or alternative downconversion material may provide only nominally less protection than thicker ones, also depending on the architecture design.

\newpage
\bibliography{apssamp}

\end{document}